\documentclass{elsart}
\usepackage{graphicx}
\usepackage{amsfonts}
\usepackage{indentfirst}
\usepackage[small]{caption2}
\usepackage{longtable}
\usepackage{epsfig}

\def\MSbar{$\overline{\mathrm{MS}}\ $}

\def\Li#1#2{{\mathrm{Li}}_{#1}\left(#2\right)}

\def\ba{\begin{eqnarray}}
\def\ea{\end{eqnarray}}
\def\DD{{\mathcal D}}
\def\dd{{\mathrm d}}

\def\fun#1#2{\lower3.6pt\vbox{\baselineskip0pt\lineskip.9pt
  \ialign{$\mathsurround=0pt#1\hfil##\hfil$\crcr#2\crcr\sim\crcr}}}
\def\order#1{{\mathcal O}\left(#1\right)}

\def\gsim{\mathrel{\raise.3ex\hbox{$>$\kern-.75em\lower1ex\hbox{$\sim$}}}}
\def\lsim{\mathrel{\raise.3ex\hbox{$<$\kern-.75em\lower1ex\hbox{$\sim$}}}}

\begin{document}
\begin{frontmatter}

\title{Next-to-leading order corrections to Bhabha scattering
in renormalization group approach (I). Soft and virtual photonic contributions.}


\author{A.B.~Arbuzov},
\author{E.S.~Scherbakova}

\address{Bogoliubov Laboratory of Theoretical Physics, \\
JINR,\ Dubna, \ 141980 \ \  Russia }

\begin{abstract}
Soft and virtual loop photonic contributions to 
the second order next-to-leading QED radiative corrections to 
Bhabha scattering are calculated with help of the renormalization
group approach. The results are in agreement with earlier 
calculations, where other methods were used. Scale dependence
and the present theoretical accuracy of Bhabha cross section 
description are discussed. 
\end{abstract}


\end{frontmatter}

\section{Introduction}

Precision theoretical predictions for the differential cross section of
Bhabha scattering are of ultimate importance for all experiments at 
electron-positron colliders. They are required for normalization 
purposes including luminosity determination, for several searches 
of new physics, and as a background contribution to many other processes 
studied at
the colliders. To provide the required accuracy of the predictions, one 
should take into account radiative corrections in the first and higher
order of perturbative QED. Certain effects of strong and weak interactions
should be included as well.

It is natural to expand the QED part of radiative corrections to Bhabha
scattering into a series in the fine structure constant $\alpha$ and in
powers of the so-called large logarithm $L=\ln(M^2/m_e^2)$, 
where $M$ is a large energy scale related to the beam energy, $M\gg m_e$. 
The terms enhanced by the large logs are known to give the bulk of the result 
in the kinematical regions of interest for experiments. 

In this paper we present the derivation of the second order next-to-leading
virtual and soft photonic radiative corrections to the cross section 
of Bhabha scattering. To get the corrections we use the renormalization
group techniques borrowed from QCD. The result is found to be in agreement 
with earlier calculations~\cite{Glover:2001ev,Penin:2005kf}. The advantage 
of our approach is its universality: in the same way one can get all the other 
remaining contributions to the radiative corrections in $\order{\alpha^2L}$. 
For the case of small angle Bhabha scattering the complete result for
the $\order{\alpha^2L}$ is known~\cite{Arbuzov:1995qd}, but for the general case of
large angle scattering terms of this order are still not systematized, while
there is a number of results for particular contributions scattered in 
the literature.

The paper is organized as follows. The notation is introduced in the next section. 
The application of the structure function approach is given in Sect.~\ref{SFA}. 
The numerical results are presented in several plots followed by a short
discussion and conclusions. In sake of completeness, some useful formulae 
are listed in the Appendix.

\section{Preliminaries}
  
Let us represent the differential Bhabha cross section as a series in $\alpha$:
\ba \label{gensig}
\dd\sigma = \dd\sigma^{\mathrm{Born}} + \dd\sigma^{(1)} + \dd\sigma^{(2)}
+ \order{\alpha^3},
\ea
where $\dd\sigma^{\mathrm{Born}}$ represents the Born level cross section, and
$\dd\sigma^{(1,2)} $ are the pure QED contributions of the first and second 
order corrections. 

The first order contribution is usually decomposed into three parts:
\ba \label{o1}
\dd \sigma^{(1)} = \dd\sigma^{\mathrm{V}} + \dd\sigma^{\mathrm{S}} 
+ \dd \sigma^{\mathrm{H}},
\ea
where superscripts "V", "S", and "H" are used to denote the virtual, soft, and 
hard  photonic corrections, respectively. 
The small parameter $\Delta$ $\ (\Delta \ll 1)$
subdivides the kinematical domain of real photon emission into soft and hard parts 
with photon energy below and above $\Delta\cdot E_{\mathrm{beam}}$, where 
$E_{\mathrm{beam}}$ is the beam energy. The one-loop contributions are well known,
see i.e. Refs.~\cite{Beenakker:1990mb,Glover:2001ev,Arbuzov:1997pj}.

In the second order we construct a similar decomposition:
\ba \label{o2}
\dd \sigma^{(2)} &=& \dd\sigma^{\mathrm{VV}} + \dd\sigma^{\mathrm{SV}} 
+ \dd\sigma^{\mathrm{SS}} 
+ \dd\sigma^{\mathrm{VH}} + \dd  \sigma^{\mathrm{SH}} + \dd \sigma^{\mathrm{HH}}, 
\ea
where the superscripts have the same meaning as in Eq.~(\ref{o1}), so that for instance
``VH'' denotes the contribution due to emission of one hard photon accompanied
by the effect of a single virtual loop. 

In this paper we don't take into account the so-called pair contributions, related
to emission of real or virtual pairs ($e^+e^-$, $\pi^+\pi^-$ {\it etc.}). 
Their numerical contribution to the observed cross section is
typically small compared to the photonic correction, see 
Ref.~\cite{Arbuzov:1995vi,Arbuzov:1995vj,Arbuzov:2001rt}.

The second order contribution $\dd\sigma^{(2)}$ is decomposed into series 
in the powers of the large logarithm.
There in particular contributions we meet terms of the orders $\order{\alpha^{2} L^{4,3,2,1,0}}$. The terms with the fourth and the third powers of $L$ will
cancel out in the sum of virtual and soft photon contributions.

Using the factorization properties of soft photon radiation, we can write 
immediately the Soft-Soft (SS) and Soft-Virtual (SV)
contributions~\cite{Arbuzov:1998du}:
\ba \label{SSSV}
\dd \sigma^{\mathrm{SS}} &=& \frac{1}{2!}(\delta^{\mathrm{S}})^2 
\dd\sigma^{\mathrm{Born}}, 
\qquad
\dd \sigma^{\mathrm{SV}} = \delta^{\mathrm{S}}\delta^{\mathrm{V}} 
\dd\sigma^{\mathrm{Born}},
\ea
where $\delta^{\mathrm{S,V}} = \dd\sigma^{\mathrm{S,V}}/\dd\sigma^{\mathrm{Born}}$
(see Appendix for explicit formulae).
Note that for the contribution of double soft photon emission, 
$\dd\sigma^{\mathrm{SS}}$,
in the formula above we apply an upper limit on the energy of each of the photons independently. The Virtual-Virtual (VV) contribution can't be received in
such a simple manner. Below we will show how to reconstruct the logarithmically
enhanced part of it using the
renormalization group techniques or in other words the electron structure 
function approach.


\section{Structure Function Approach \label{SFA}}

The structure (fragmentation) function approach, widely used in QCD can be
applied to QED 
problems~\cite{Kuraev:1985hb,Berends:1987ab,Skrzypek:1992vk,Arbuzov:1999cq,Arbuzov:2002cn}. 
With help of it we can analytically find the most
important contributions reinforced by the large logarithm $L$, since they 
can be treated as electron mass singularities.

We are going to drop the pair contributions, so we need here the pure photonic 
part of the non-singlet structure (fragmentation) functions for the initial (final)
state corrections. These functions describe the
probability to find a massless (massive) electron with energy fraction $z$ 
in the given massive (massless) electron.   
In our case with the next-to-leading accuracy we have
\ba  \label{Dee}
\DD_{ee}^{\mathrm{str,frg}} (z) &=& \delta(1-z)
+ \frac{\alpha}{2\pi}d_1(z,\mu_0,m_e)
+ \frac{\alpha}{2\pi}LP^{(0)}_{ee}(z)
 \nonumber \\
&+& \biggl(\frac{\alpha}{2\pi}\biggr)^2
\biggl(\frac{1}{2}L^2P^{(0)}_{ee}\otimes P^{(0)}_{ee}(z)
+ LP^{(0)}_{ee}\otimes d_1(z,\mu_0,m_e)
\\ \nonumber 
&+& LP^{(1,\gamma){\mathrm{str,frg}}}_{ee}(z) \biggr)
+ \order{\alpha^2 L^{0}, \alpha^3},
\ea
where the superscripts ``str'' and ``frg'' are used to mark the structure and 
fragmentation functions, respectively. The difference between the functions appear
only due to the difference in the next-to-order splitting functions $P^{(1,\gamma)}$.
The modified minimal subtraction scheme \MSbar is used.
We have chosen the factorizations scale equal to $M$, and the renormalization scale
$\mu_0$ will be taken equal to $m_e$. 
More details on application of the approach to calculation
of second order next-to-leading QED corrections can be found in 
Refs.~\cite{Berends:1987ab,Arbuzov:2002cn}. 

The function $d_1$ corresponds to next-to-leading order initial condition
in the \MSbar scheme for iterative solutions of the QED DGLAP equation. It reads
\ba
d_1(z,\mu_0,m_e) &=& \biggl[ \frac{1+z^2}{1-z}
\biggl( \ln\frac{\mu_0^2}{m_e^2} - 2\ln(1-z) - 1 \biggr)
\biggr]_+\, ,
\ea
where the usual plus prescription is exploited to regularize the singularity at
$z\to 1$. The relevant lowest order splitting function is
\ba
P_{ee}^{(0)}(z)= \biggl[\frac{1+z^2}{1-z}\biggr]_+.
\ea
The next-to-leading order functions $P^{(1,\gamma)}_{ee}$ is provided by 
the set of Feynman diagrams with pure photonic corrections,
\ba
P^{(1,\gamma){\mathrm{frg}}}_{ee}(z) &=&
\delta(1-z)\biggl( \frac{3}{8} - 3\zeta(2) + 6\zeta(3) \biggr)
+ \frac{1+z^2}{1-z}\biggl( 2\ln z \ln(1-z)
- 2\ln^2z 
\nonumber \\ 
&-& 2\Li{2}{1-z} \biggr) + \frac{1}{2}(1+z)\ln^2z
+ 2z\ln z - 3z + 2,
\\ \nonumber 
P^{(1,\gamma){\mathrm{str}}}_{ee}(z) &=& 
\delta(1-z)\biggl( \frac{3}{8} - 3\zeta(2) + 6\zeta(3) \biggr)
+ \frac{1+z^2}{1-z}\biggl( - 2\ln z \ln(1-z)
+ \ln^2z 
\nonumber \\ 
&+& 2\Li{2}{1-z} \biggr) - \frac{1}{2}(1+z)\ln^2z
+ 2\ln z + 3 - 2z.
\ea
Note that in these splitting functions the parts proportional to the delta-function are 
equal, and the remaining functions are not singular at $z\to 1$.
Therefore, since at the present moment we are interested only in the soft and virtual
part of the corrections, we can forget about the difference and drop the superscripts
distinguishing the structure and fragmentation functions. 

The master formula describing the radiatively corrected Bhabha cross section in
the structure function approach reads~\cite{Arbuzov:1997pj}:
\ba \label{master}
\dd \sigma &=& \int^{1}_{\bar{z_1}} \dd z_1 \int^{1}_{\bar{z_2}} \dd z_2 
          \DD^{\mathrm{str}}_{ee} (z_1) \DD^{\mathrm{str}}_{ee} (z_2)
		\left( \dd \sigma^{\mathrm{Born}} (z_1,z_2) + \dd \bar{\sigma}^{(1)} (z_1,z_2) + \order{\alpha^2L^0}  \right) 
\nonumber \\
&\times& \int^{1}_{\bar{y_1}} \frac{\dd y_1}{Y_1} \int^{1}_{\bar{y_2}} \frac{\dd y_2}{Y_2}
\DD^{\mathrm{frg}}_{ee} (\frac{y_1}{Y_1}) \DD^{\mathrm{frg}}_{ee} (\frac{y_2}{Y_2}),
\ea
where $\dd\bar\sigma^{(1)}$ is the $\order{\alpha}$ correction to the massless Bhabha scattering,
calculated using the \MSbar scheme to subtract the lepton mass singularities.  
Energy fractions of incoming partons are $z_{1,2}$, and 
$Y_{1,2}$ are the energy fractions of the outcoming electron and positron.

Here we are interested in the contributions due to virtual and soft photons, so 
all the four integrals will have the same lower limit being equal to $1-\Delta$.
First we can perform convolution of the four structure functions entering
Eq.~(\ref{master}) with each other:
\ba
D^{\otimes 4} (z) &=& \biggl(  \delta(1-z) + \frac{\alpha}{2 \pi} d_1(z) + 
\frac{\alpha}{2 \pi} L P^{(0)}(z) \biggr)^{\otimes 4} 
 \nonumber \\
&=& \delta(1-z) 
+ 4 \frac{\alpha}{2 \pi}\biggl( d_1(z) + L P^{(0)}(z) \biggr) 
+ 4\biggl(\frac{\alpha}{2 \pi}\biggr)^2 \biggl( 2L^2(P^{(0)}(z))^{\otimes 2}
\nonumber \\
&+& 4L \,d_1(z)\otimes P^{(0)}(z) + L P^{(1)}(z)\biggr)
+ \order{\alpha^2L^0,\alpha^3},
\ea						 
with a short notation for multiple convolution operation: 
$(A)^{\otimes n} \equiv \underbrace{A\otimes A \otimes A \cdots A}_n$.
If $z=1-\Delta$ and $\Delta\ll 1$, this function gives the probability to 
find such a situation where one looses in total due to photon emission 
$\Delta E_{\mathrm{beam}}$ from the total energy of the process under consideration
(the center-of-mass reference frame is used throughout the paper).

Let us fix now the factorization scale $M=\sqrt{s}$ and define 
\ba
L_s\ \equiv\ \ln\frac{s}{m_e^2}\, , \qquad s = 4E_{\mathrm{beam}}^2.
\ea
Later on we will consider another choice of the scale.

Convolution of the function found above with the Born part of the kernel cross section 
gives us the corresponding part
to the cross section (with the upper limit on the energy lost):
\ba \label{d4sig0}
&& \int^1_{1 - \Delta} \DD^{\otimes 4}(z) \dd \sigma^{\mathrm{Born}}(z) \dd z =  
\dd\sigma^{\mathrm{Born}} \Biggl\{ 1 + \frac{\alpha}{2 \pi} 
\biggl[  4L_s\biggl( 2 \ln\Delta +\frac{3}{2}\biggr) + \order{L_s^0} \biggr]  
\nonumber \\ &&\qquad 
+ \biggl(\frac{\alpha}{2 \pi}\biggr)^2 \biggl[ 
  8 L_s^2 \left(P^{(0)}\right)^{\otimes 2}_{\Delta}
+ 16 L_s (P^0 \otimes d_1)_{\Delta} + 4 L_s P^{(1)}_{\Delta}  \biggr]
+ \order{\alpha^2L_s^0,\alpha^3}
\Biggr\},
\ea
where we used subscript $\Delta$ to specify the the so-called $\Delta$-part of the
corresponding function (see i.e. Refs.~\cite{Skrzypek:1992vk,Arbuzov:1999cq}). 

Convolution with the $\dd\bar{\sigma}^{(1)}$ is more complicated since the latter is
a non-trivial function of $z$. But we restricted ourselves to consider only the terms
reinforced by the large logarithm. Therefore we need to compute only the following
part:
\ba \label{sig1P}
4\frac{\alpha}{2\pi}L_s\int^1_{1 - \Delta}\dd z [P^{(0)} \otimes \dd \bar{\sigma}^{(1)}](z).
\ea
Using the techniques of dealing with the singular functions regularized by introduction
of the $\Delta$ and $\Theta$ parts~\cite{Skrzypek:1992vk,Arbuzov:1999cq,Arbuzov:2003ed} 
we cast it into the following form:
\ba \label{sigP}
4\frac{\alpha}{2\pi}\int^1_{1 - \Delta}\dd y \int_0^1\frac{\dd z}{z}\; 
L_sP^{(0)}\left(\frac{y}{z}\right)  
\dd \bar{\sigma}^{(1)}(z) = 4\frac{\alpha}{2\pi}\, 
\dd \bar{\sigma}^{(1)}_\Delta L_s P^{(0)}_{\Delta}
+ 4\frac{\alpha}{2\pi}\, L_s \delta (\bar{\sigma}^{(1)}),
\ea
where
\ba
\dd\bar{\sigma}^{(1)}_\Delta &=& \int^1_{1 - \Delta}\dd\bar{\sigma}^{(1)}(z)\dd z
= \dd\sigma^{\mathrm{V}} + \dd\sigma^{\mathrm{S}} 
- \dd\sigma^{\mathrm{Born}}\frac{\alpha}{2\pi}\, 4 L_sP^{(0)}_\Delta,
\nonumber \\
\delta (\bar{\sigma}^{(1)}) &=& \int_{1-\Delta}^1 \dd \bar{\sigma}^{(1)}(z)\; 
       2\; \ln\biggl(1-\frac{1-z}{\Delta}\biggr) \dd z. 
\ea
Let us note that the last non-trivial term the function under the integral above
is not singular for $z\to 1$. So the virtual loop contribution to the function 
$\dd\bar{\sigma}^{(1)}$ doesn't contribute to the result of the integral. And we can 
put the upper limit of the integral over $z$ to be equal to $1-\Delta_1$, 
$\ \Delta_1 \ll \Delta \ll 1$.
Therefore we need to consider only the soft photon part of the function:
\ba
\!\!\!\!
\frac{\dd\bar{\sigma}^{(1,{\mathrm{S}})}(z)}{\dd z}  &=& 
\dd\sigma^{\mathrm{Born}}\Biggl\{
- \frac{\alpha}{4 \pi^2}
\frac{\dd |\vec{k}|}{\dd z}
\frac{|\vec{k}|^2
}{\sqrt{|\vec{k}|^2 + \lambda^2}}
 \left[\int_0^{2 \pi}\! \dd\phi \int_{-1}^1\! \dd c 
\left(\frac{p_{e^+}}{p_{e^+} k}  - \frac{p_e}{p_e k} 
+ \frac{p'_e}{p'_e k} - \frac{p'_{e^+}}{p'_{e^+}k} \right)^2
		\right] 
\nonumber \\
&-& \frac{\alpha}{2 \pi}\, 4 
\left[ d_1(z) + L_s P^{(0)}(z)\right]\Biggr\},
\ea
where $1-z = |\vec{k}|/E_{\mathrm{beam}}$;
$\lambda$ is a fictitious photon mass, $\lambda \ll m_e$; $p_e$ $(p_{e^+})$ 
is momentum of incoming electron 
(positron),  $p'_e$, $p'_{e^+}$, and $k$ are the momenta of the outgoing electron, positron, 
and the photon.
Using the standard techniques of calculations of soft photon contributions
we get
\ba \label{dsig1}
\frac{\delta (\bar{\sigma}^{(1)})}{\dd\sigma^{\mathrm{Born}}} 
&=& \frac{1}{\dd\sigma^{\mathrm{Born}}} \int^{1-\Delta_1}_{1-\Delta}\dd z
\frac{\dd\bar{\sigma}^{(1,{\mathrm{S}})}(z)}{\dd z}\, 
2 \ln\left(1-\frac{1-z}{\Delta}\right)
\nonumber \\ 
&=& 4\frac{\alpha}{2\pi} \int^{1-\Delta}_{1-\Delta_1} \frac{\dd z}{1-z} 
\biggl[8 \ln (1-z) + 4 \ln \frac{1-c}{2} 
- 4 \ln\frac{1+c}{2} \biggr] \ln\left(1-\frac{1-z}{\Delta}\right)
\nonumber \\ 
&=&  4\frac{\alpha}{2\pi} \biggl[ - 4 \zeta(2) \ln\frac{1-c}{1+c} + 
   8 \biggl( \zeta(3) - \zeta(2)\ln\Delta \biggr)\biggr],
\ea
where $c$ is the cosine of the electron scattering angle, $c=\cos\widehat{\vec{p}_e\vec{p}'}_{e}$.

Summing up the contributions in Eq.~(\ref{sigP}) 
and then summing up Eqs.~(\ref{SSSV},\ref{d4sig0},\ref{sigP}) we 
receive the leading and next-to-leading second order contributions to quasi-elastic
Bhabha cross section, where the total energy loss (due to soft photon emission) is 
limited by $\Delta E_{\mathrm{beam}}$. 
It is useful to describe also the case when the energies of the soft photons 
(if they are two) are limited independently. 
The transition between the two cases was derived
in Ref.~\cite{Arbuzov:1998du}. Applying it we get the final result:
\ba \label{result}
\dd\sigma^{\mathrm{VV}}&+&\dd\sigma^{\mathrm{SV}} 
+ \dd\sigma^{\mathrm{SS}} = \left(\frac{\alpha}{2\pi}\right)^2
\dd\sigma^{\mathrm{Born}}\biggl\{
L_s^2\biggl[ 32\ln^2\Delta +48 \ln\Delta  +18 \biggr]  \nonumber\\
&+& L_s\biggl[ 
 64 \biggl(  \ln\left(\frac{x}{1-x}\right) - 1 \biggr)
 \ln^2\Delta  \nonumber\\
 &+& 16 \biggl( 2\,\Li{2}{1-x} - 2\,\Li{2}{x}+3\ln\left(\frac{x}{1-x}\right) 
  + f(x) - 7 -\frac{2}{3}\pi^2 \biggr) 
\ln\Delta    \nonumber\\
&+& 24\,\Li{2}{1-x}-24\,\Li{2}{x} + 12 f(x)
 + 24 \zeta(3) -\frac{93}{2} - 10 \pi^2
\biggr]  \nonumber \\
&+& 4\delta_0^{(2)} + \order{\frac{m_e^2}{s}}
\biggr\},\qquad
x \equiv \frac{1-c}{2}\, , 
\ea
where the function $f(x)$ is given in Appendix.
In the formula above besides the logarithmically enhanced terms derived here, 
we included also the known contribution without the large logs, 
$\delta_0^{(2)}$, which is given by Eq.~(3) from Ref.~\cite{Penin:2005kf}.

\section{Numerical Results and Conclusions}

In this way we received the photonic part of the second order next-to-leading
logarithmic contribution to Bhabha cross section. The result agree with earlier
calculations by means of different methods. Our approach allows to get all
the next-to-leading contributions systematically. It can be applied to any kind of 
a process, where one has to look for the radiative corrections enhanced by 
large logarithms. In particular, we applied the same approach to the description of 
the contribution of real photon emission to Bhabha scattering [to be described elsewhere]. 

Let us compare the numerical values of the leading, next-to-leading, 
and next-to-next-to-leading corrections for two choices of the factorization
scale: $M=\sqrt{s}$, which has been used in Refs.~\cite{Glover:2001ev,Penin:2005kf}, 
and $M=\sqrt{-t}$, which has been advocated in Ref.~\cite{Arbuzov:1995qd}.
Since we have the complete answer~(\ref{result}), we can easily choose any other 
the factorization scale by changing the argument of the large logarithm, while the 
total sum is kept unchanged. For this purpose we use the relations 
\ba
t = - s\frac{(1-c)}{2} = - xs, \qquad L_s = L_t - \ln x, \qquad 
L_t\equiv\ln\frac{-t}{m_e^2}\, .
\ea

In Figures~1---5 we show the values of the second order soft and virtual photonic
radiative corrections in different approximations with respect 
to the power of the large logarithm. Values of the corrections are given in terms
of $10^{-3}\cdot\dd\sigma^{\mathrm{Born}}$. In particular, $r^{(2)}_{2,1,0}$
represent the leading $\order{\alpha^2L^2}$, the next-to-leading $\order{\alpha^2L^1}$,
and the next-to-next-to-leading $\order{\alpha^2L^0}$ relative contributions 
to the cross section, respectively. Since the dependence on the parameter
$\Delta$ should disappear in the sum of the virtual and soft corrections with the
remaining three contributions (see Eq.~(\ref{o2})), we put $\Delta=1$. In this way 
we receive only an estimate of the magnitude and the relative size of the corrections
in different approximations. Nevertheless this evaluation helps us to get
an idea about the size of the unknown second order contributions and to estimate the
theoretical uncertainties.

\begin{figure}[ht]
\begin{center}
\includegraphics*[width=7cm,height=7.6cm,keepaspectratio,angle=270]{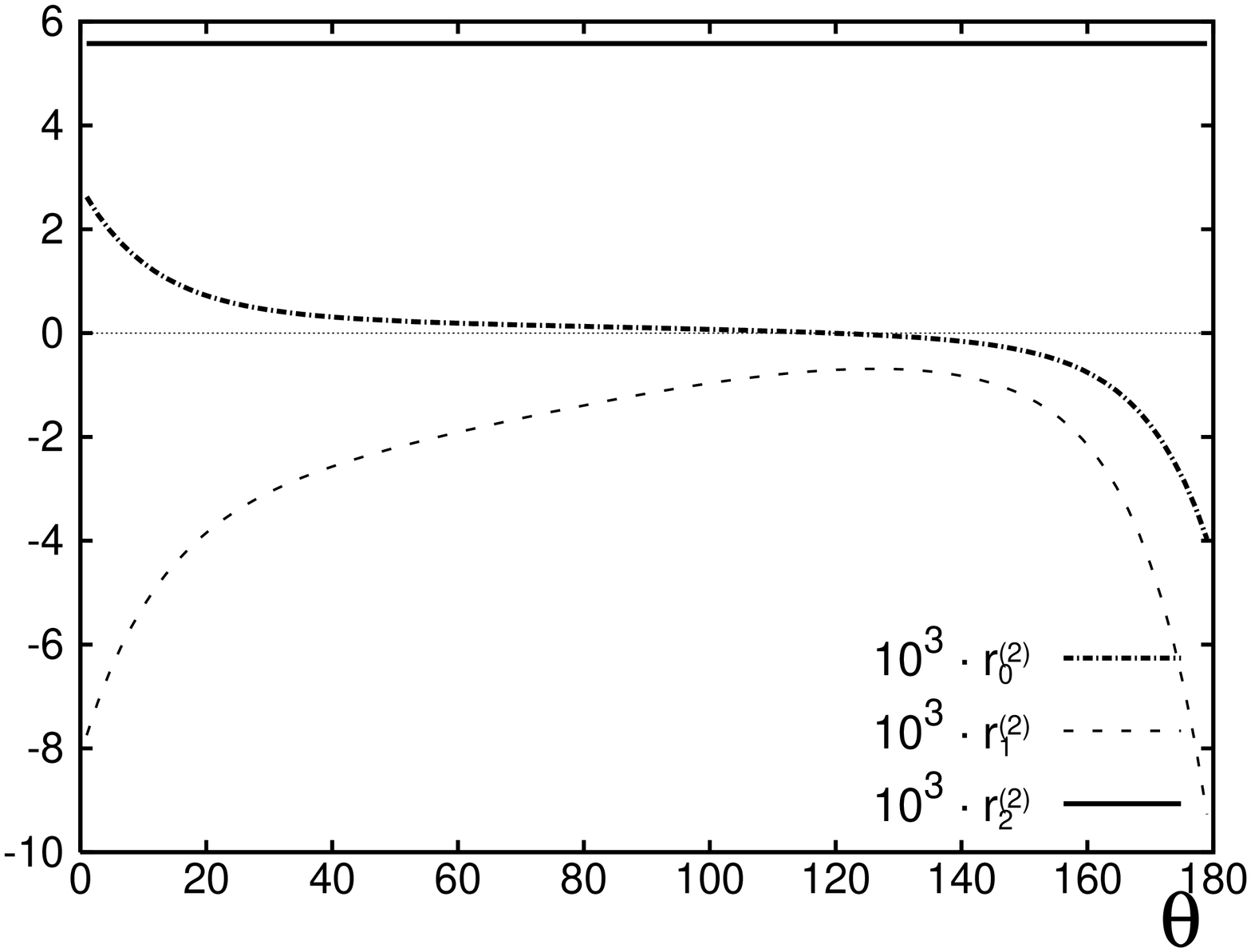}
\includegraphics*[width=7cm,height=7.6cm,keepaspectratio,angle=270]{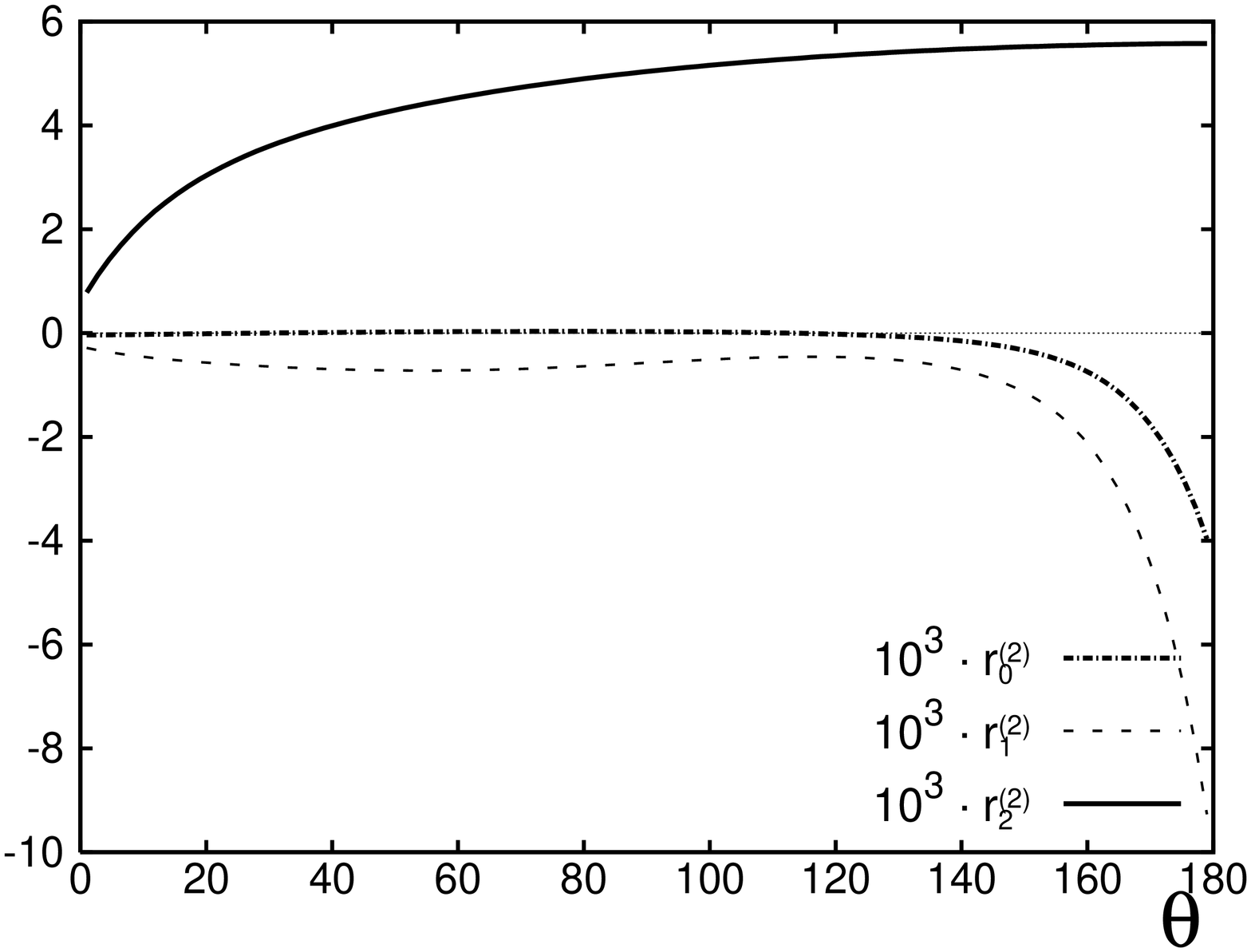}
\end{center}
\caption{Soft and virtual second order photonic radiative corrections {\it versus} 
the scattering angle in degrees for $\Delta=1$,
$\ \sqrt{s}$=1 GeV; $M=\sqrt{s}$ on the left side and $M=\sqrt{-t}$ on the right side. 
}
\label{pi1ST}
\end{figure}
Fig.~\ref{pi1ST} shows the size of the corrections under consideration relevant
for the studies at electron-positron colliders of moderate energy (VEPP-2M, BEPS {\it etc.}).

\begin{figure}[ht]
\begin{center}
\includegraphics*[width=7cm,height=7.6cm,keepaspectratio,angle=270]{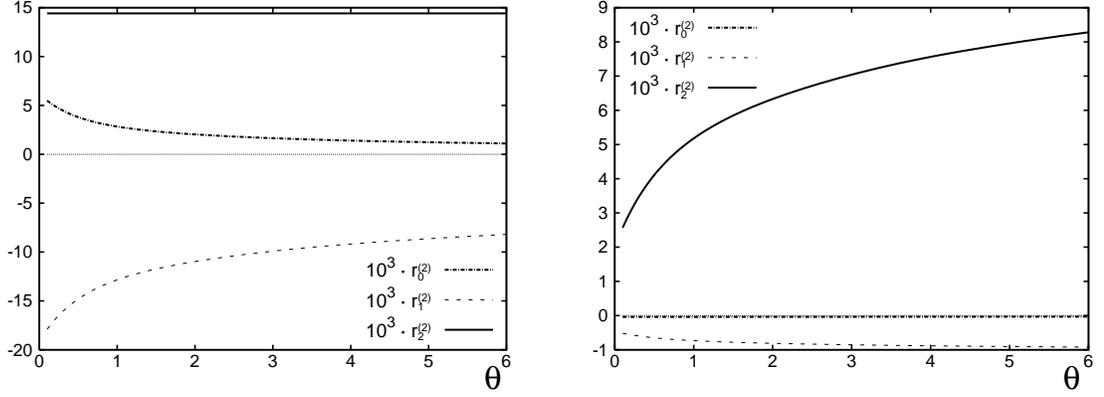}
\includegraphics*[width=7cm,height=7.6cm,keepaspectratio,angle=270]{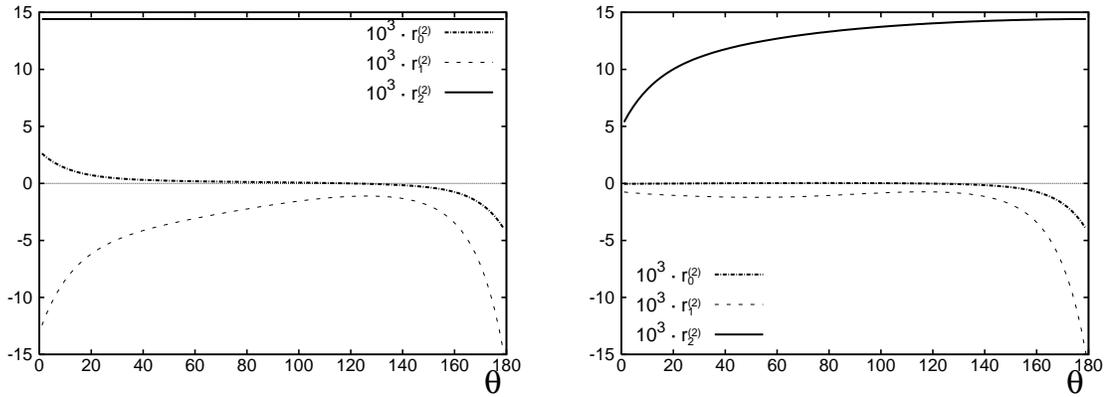}
\end{center}
\caption{Soft and virtual second order photonic radiative corrections {\it versus} 
the scattering angle in degrees for $\Delta=1$, 
$\ \sqrt{s}$=100 GeV; $M=\sqrt{s}$ on the left side and $M=\sqrt{-t}$ on the right side. 
}
\label{100ST}
\end{figure}

\begin{figure}[ht]
\begin{center}
\includegraphics*[width=7cm,height=7.6cm,keepaspectratio,angle=270]{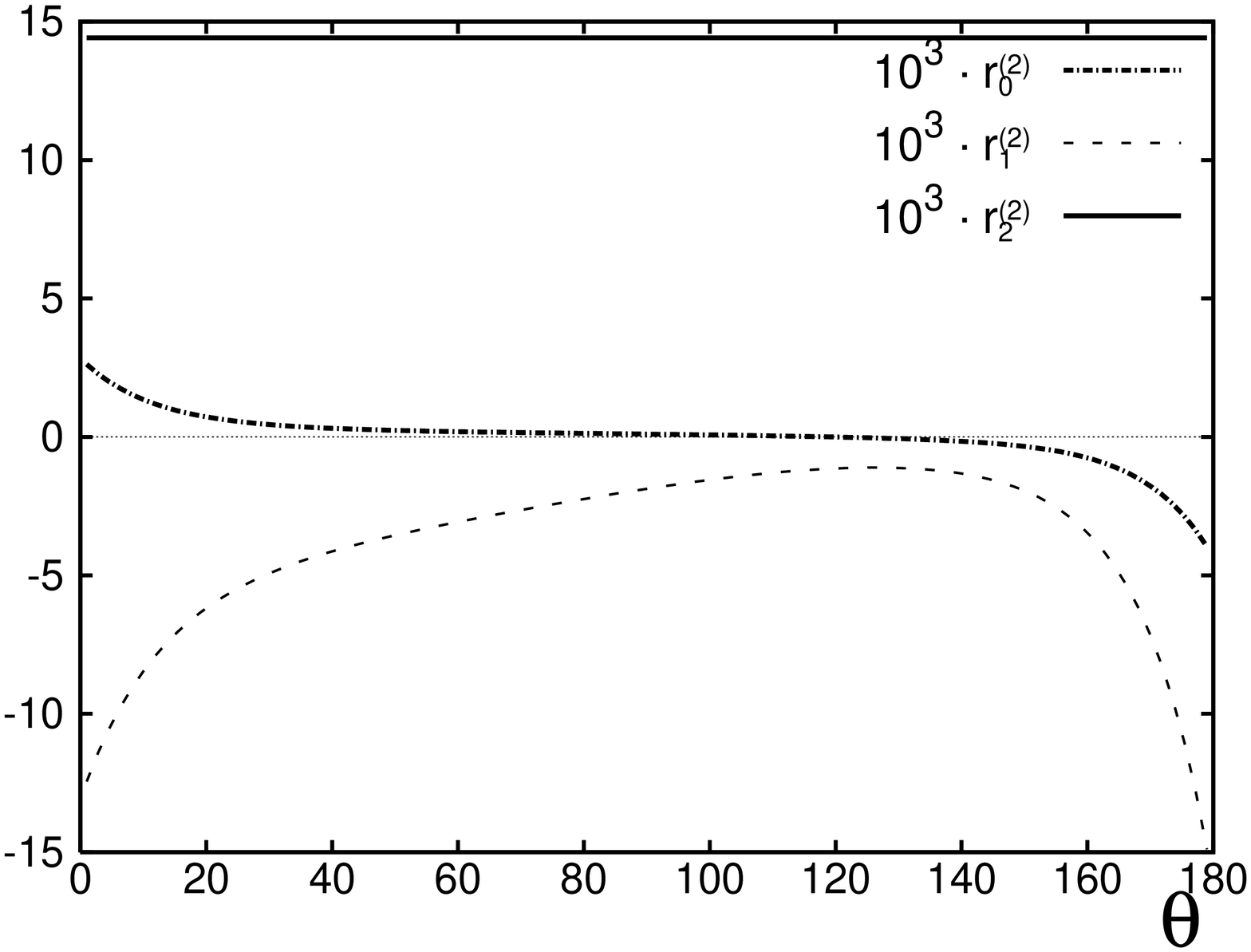}
\includegraphics*[width=7cm,height=7.6cm,keepaspectratio,angle=270]{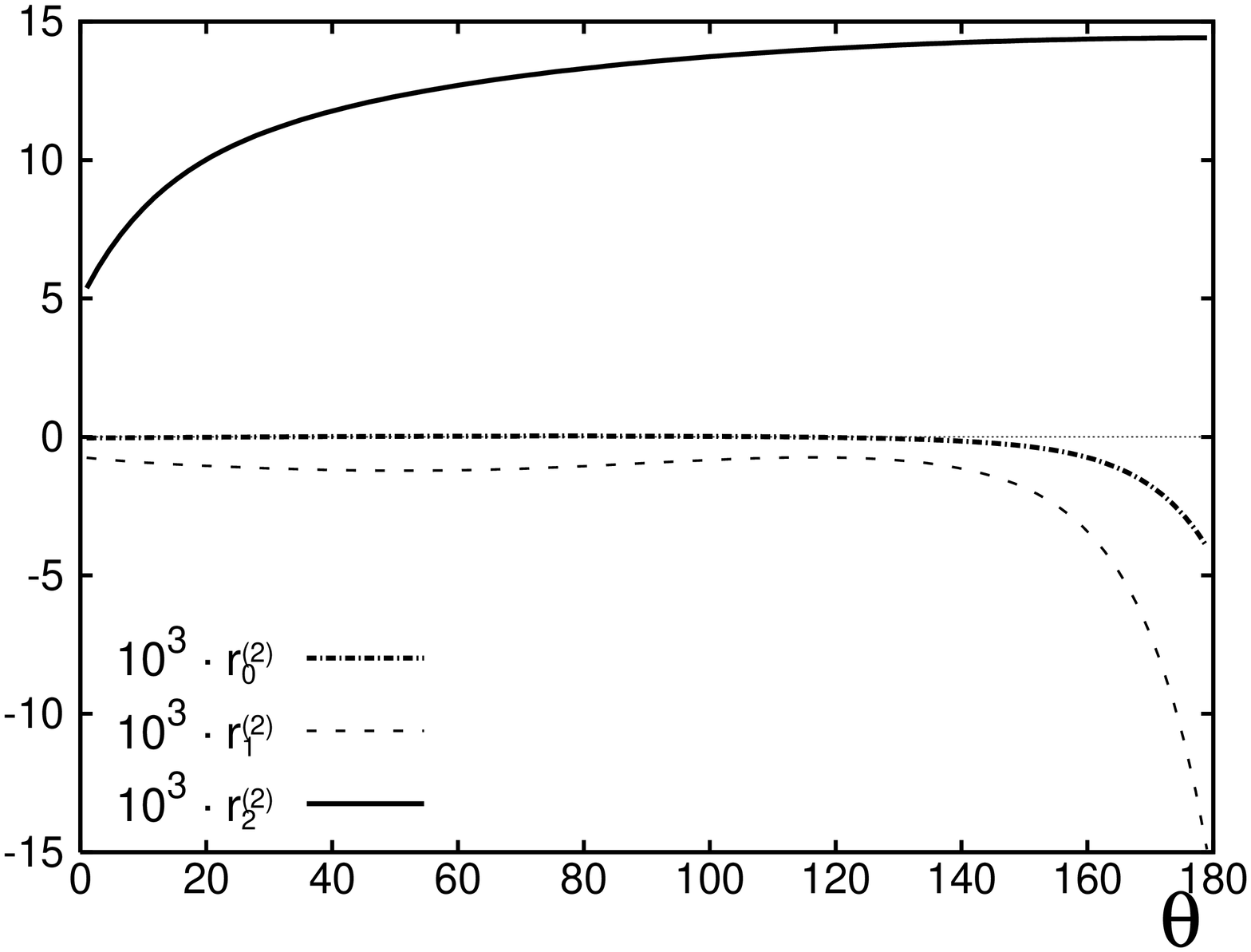}
\end{center}
\caption{Soft and virtual second order photonic radiative corrections {\it versus} 
the scattering angle in degrees for $\Delta=1$, 
 $\ \sqrt{s}$=100 GeV; $M=\sqrt{s}$ on the left side and $M=\sqrt{-t}$ on the right side. 
}
\label{pi100ST}
\end{figure}
Fig.~\ref{100ST} and Fig.~\ref{pi100ST} gives us results for the small and large angle
Bhabha scattering at LEP/SLC, respectively. We checked that for $\sqrt{s}$=200~GeV
the plots are very close the the ones shown for $\sqrt{s}$=100~GeV.

%

\begin{figure}[ht]
\begin{center}
\includegraphics*[width=7cm,height=7.6cm,keepaspectratio,angle=270]{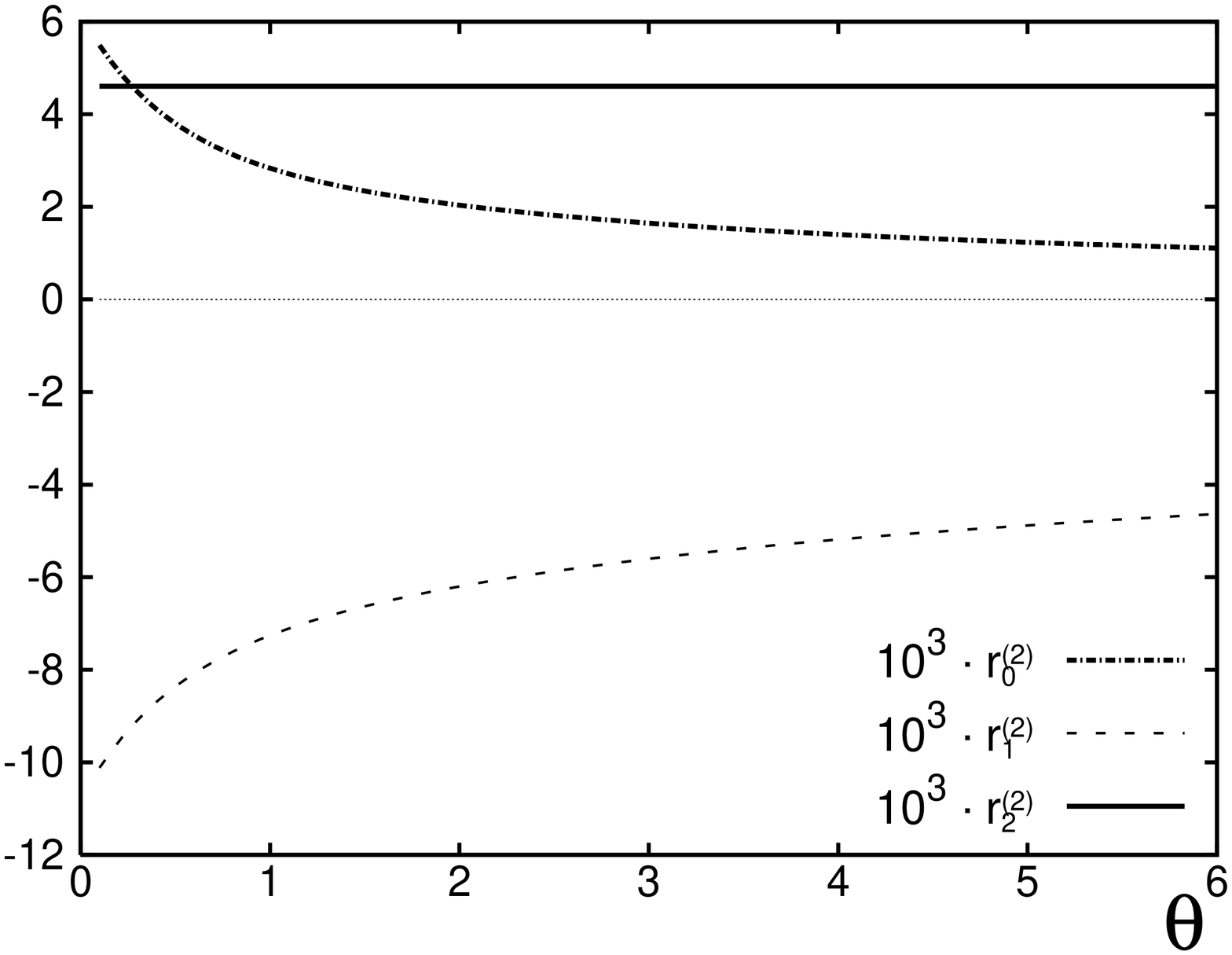}
\includegraphics*[width=7cm,height=7.6cm,keepaspectratio,angle=270]{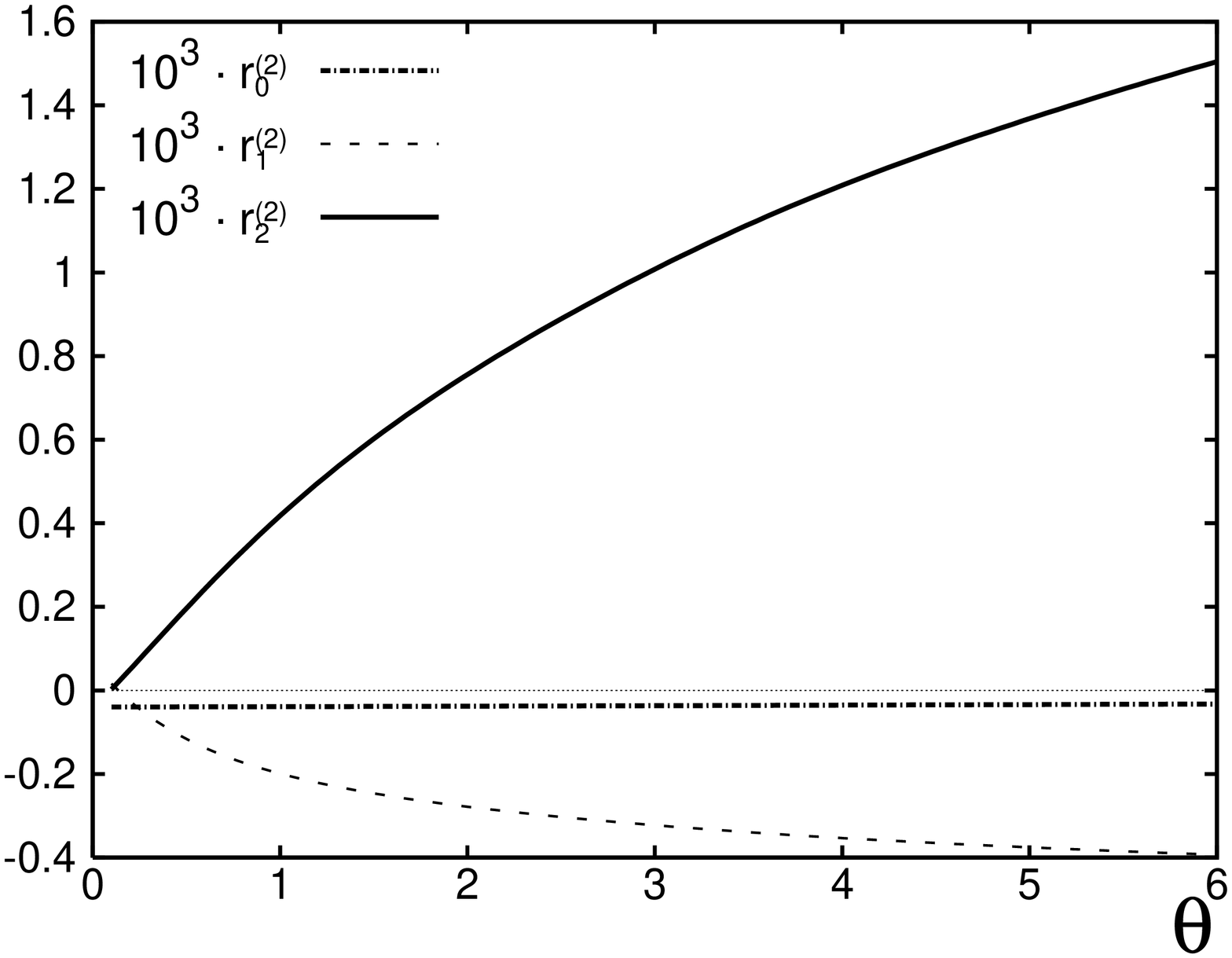}
\end{center}
\caption{Soft and virtual second order photonic radiative corrections {\it versus} 
the scattering angle in degrees for $\Delta=1$, 
$\ \sqrt{s}$=500 GeV; $M=\sqrt{s}$ on the left side and $M=\sqrt{-t}$ on the right side. 
}
\label{500ST}
\end{figure}

\begin{figure}[ht]
\begin{center}
\includegraphics*[width=7cm,height=7.6cm,keepaspectratio,angle=270]{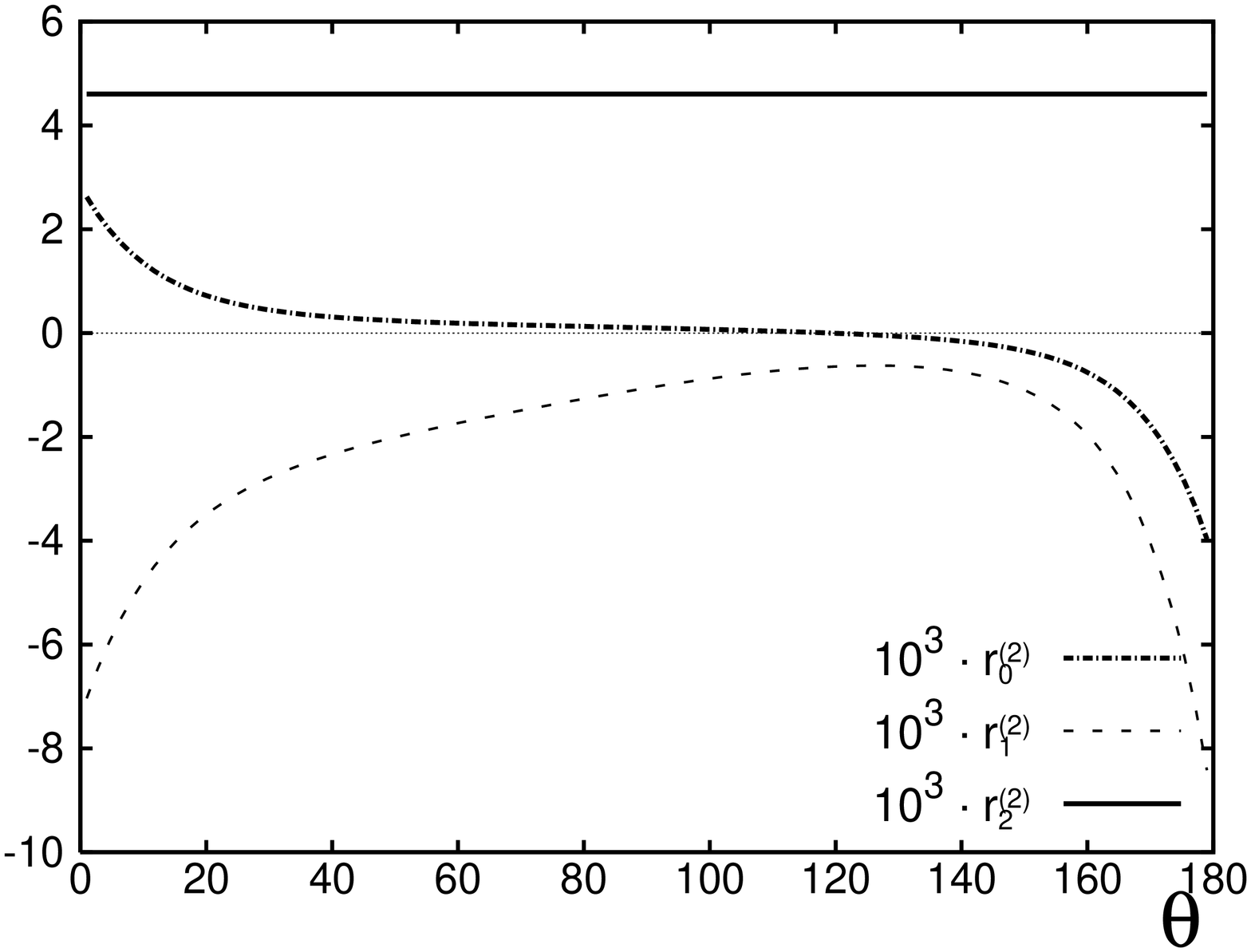}
\includegraphics*[width=7cm,height=7.6cm,keepaspectratio,angle=270]{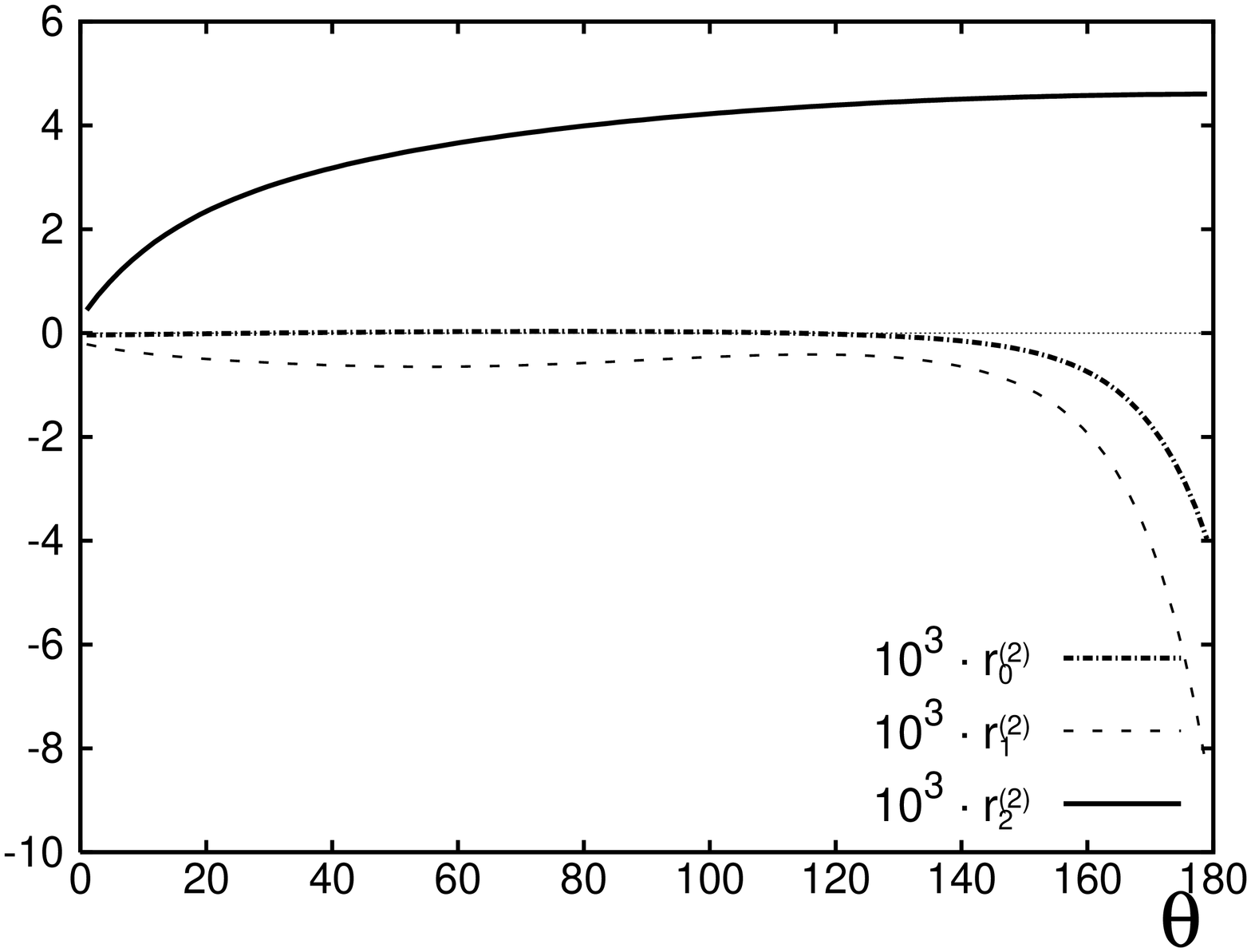}
\end{center}
\caption{Soft and virtual second order photonic radiative corrections {\it versus} 
the scattering angle in degrees for $\Delta=1$, 
$\ \sqrt{s}$=500 GeV; $M=\sqrt{s}$ on the left side and $M=\sqrt{-t}$ on the right side. 
}
\label{pi500ST}
\end{figure}

Figures~\ref{500ST} and \ref{pi500ST} give as and idea, how do the corrections behave 
at higher energies, which can be reached at a future linear collider.

Looking at the plots representing the contributions of different powers
in $L$, we conclude that with the proper choice of
the factorization scale $M=\sqrt{-t}$, the magnitude of the non-logarithmic corrections 
is below $1\cdot 10^{-4}\%$, everywhere except the region of very large
scattering angles $(\theta \gsim 160^{\circ})$. That region requires a special treatment, 
and it doesn't seem to be of interest for the experiments. Note
that the estimate of the size of the non-logarithmic second order
corrections agree with the one made earlier in Ref.~\cite{Arbuzov:1995qd}. At the same 
moment it is clear that to reach the $1\cdot 10^{-4}\%$ level in the precision
of theoretical description of Bhabha scattering we should take into account the
complete $\order{\alpha^2L^0}$ calculations including the effects of virtual and
real corrections due to pairs and photons.

Taking the result of the present study and the $\order{\alpha^2L}$ results of 
papers~\cite{Arbuzov:1995vi,Arbuzov:1995vj},
where the pair corrections were evaluated, and of Refs.~\cite{Arbuzov:1996qb,Arbuzov:1998ax}, where
real photon radiation was taken into account, 
we arrive at the complete result for the second order next-to-leading radiative
corrections to Bhabha scattering. The results are valid both for the small
and large angle scattering. To apply the results to data analysis of modern
and future experiments at electron-positron colliders, we are going to implement 
them into the Monte Carlo event generators {\tt LABSMC}~\cite{Arbuzov:1999db} and 
{\tt SAMBHA}~\cite{Arbuzov:2004wp} for large and small angle scattering, respectively.


\ack{
We are grateful to L.~Trentadue and A.~Penin for discussions.
This work was supported by RFBR grant 04-02-17192.
One of us (A.A.) thanks also the grant of the President RF
(Scientific Schools 2027.2003.2).
}


\section*{Appendix. Explicit Formulae for the Corrections}
\vskip 20.0pt
\setcounter{equation}{0}
\renewcommand{\theequation}{A.\arabic{equation}}

The differential Born level cross section reads
\ba
\frac{\dd\sigma^{\mathrm{Born}}}{\dd c} = \frac{\alpha^2}{4s}
\left(\frac{3+c^2}{1-c}\right)^2.
\ea

The one-loop virtual and soft corrections read
\ba
\label{eq:dv}
\dd\sigma^{\mathrm{V}}&=&\dd\sigma^{\mathrm{Born}}\frac{\alpha}{\pi}\biggl\{
4\ln\frac{m_e}{\lambda}
\left(1-L_s+\ln\left(\frac{1-x}{x}\right)\right)
-L_s^2+2L_s\ln\left(\frac{1-x}{x}\right)-\ln^2(x) \nonumber \\ 
&+& \ln^2(1-x) + 3L_s - 4 + f(x) \biggr\},
\\
\dd\sigma^{\mathrm{S}}&=&\dd\sigma^{\mathrm{Born}}\frac{\alpha}{\pi}\biggl\{
4\ln\left(\frac{m_e\Delta}{\lambda}\right)
\left(L_s-1+\ln\left(\frac{x}{1-x}\right)\right)
+L_s^2+2L_s\ln\left(\frac{x}{1-x}\right) \nonumber \\ 
&+&\ln^2(x) - \ln^2(1-x)-\frac{2\pi^2}{3} + 2\Li{2}{1-x} - 2\Li{2}{x} \biggr\}.
\ea

\ba
f(x)&=&(1-x+x^2)^{-2}\biggl[
\frac{\pi^2}{12}( 4 - 8 x + 27 x^2 - 26 x^3 + 16 x^4)\nonumber \\
&& +\frac{1}{2}(-2+5x-7x^2+5x^3-2x^4)\ln^2(1-x)
+\frac{1}{4}x(3-x-3x^2+4x^3)\ln^2(x)\nonumber \\
&&
+\frac{1}{2}(6 - 8 x + 9 x^2 - 3 x^3)\ln(x)
-\frac{1}{2}x(1+x^2)\ln(1-x)\nonumber \\
&&
+\frac{1}{2}(4-8x+7x^2-2x^3)\ln(x)\ln(1-x) \biggr].
\ea

The dilogarithm and the Riemann zeta-function are defined as usually:
\ba
\Li{2}{x} = \int_0^1\dd y \frac{\ln(1-xy)}{y}\,, \qquad
\zeta(n) = \sum\limits_{n=1}^{\infty}\frac{1}{n^l}\, .
\ea


\end{document}